\documentclass[twocolumn,aps,prb,superscriptaddress,floatfix]{revtex4}

\begin{document}
\title{Hubbard model vs. Kondo model: Strong coupling limit}

\author{Ilya Ivantsov}
\affiliation{Bogoliubov Laboratory of Theoretical Physics, Joint
Institute for Nuclear Research, Dubna, Russia}
\author{Alvaro Ferraz}
\affiliation{International Institute of Physics - UFRN,
Department of Experimental and Theoretical Physics - UFRN, Natal, Brazil}
\author{ Evgenii Kochetov}
\affiliation{Bogoliubov Laboratory of Theoretical Physics, Joint
Institute for Nuclear Research, Dubna, Russia}

\begin{abstract}
The
Hubbard model and the lattice Kondo model are shown to become identical in the strong-coupling limit. A departure from the strong-coupling regime produces distinct theories, however:
the relevant perturbation terms give rise to either short-range
spin exchange $(t-J$ model) or short-range charge exchange ($t-V$ model), respectively.

\end{abstract}
\maketitle


A better understanding of strongly correlated electrons is crucial to
address strongly coupled electronic systems in general and, in particular,
the so-called Mott physics in the pseudogap phase of high temperature
superconductors. There is no question that at very light dopings , in the
vicinity of the Mott transition, the strong antiferromagnetic (AF)
correlations take over and are dominant over all other effects. However at
larger dopings but still inside this pseudogap phase there are several
other well observed features such as Fermi surface reconstructions and the
presence of charge density waves (CDW) that are indicative that other
important effects may also be at play. To find new clues for these
different scenarios we compare the Hubbard and the lattice Kondo models. We
show that they are identical to each other at strong coupling and describe
different physics as soon as we depart from such a regime.

The Hubbard model Hamiltonian reads
\begin{equation}
H_{U} = -t\sum_{ij\sigma}
c_{i\sigma}^{\dagger} c_{j\sigma} + U\sum_i
n_{i\uparrow}n_{i\downarrow},
\label{1.0}\end{equation}
where $c_{i\sigma}$ is the annihilation operator of electron on a site $i$ with spin projection $\sigma=\uparrow,\downarrow$, $t$ is the hoping amplitude between the nearest neighbour (nn) sites. The on-site electron number operator $n_i=\sum_{\sigma}c^{\dagger}_{i\sigma}c_{i\sigma}$, and $U$ is the on-site Coulomb repulsion.

Consider the large-$U$ limit, $U/t\gg 1$. Since the on-site operator $n_{i\uparrow}n_{i\downarrow}$ possesses only non-negative eigenvalues,
the large $U$ coupling enforces the no double occupancy (NDO) constraint {\it locally}, i.e., $n_{i\uparrow}n_{i\downarrow}=0.$ In this limit,
\begin{equation}
H_{U= \infty}=
-t\sum_{ij,\sigma}\tilde c^{\dagger}_{i\sigma}\tilde c_{j\sigma}.
\label{1.1}\end{equation}
The projected electron operators
\begin{equation}
\tilde c^{\dagger}_{i\sigma}=c^{\dagger}_{i\sigma}(1-n_{i-\sigma}),\,\,n_{i\uparrow}n_{i\downarrow}=0,
\label{1.2}\end{equation}
are isomorphic to the generators of the Hubbard ($su(2|1$)) algebra \cite{wieg} spanned by the operators $X^{pq}=|p\rangle\langle q|,$ where
$p,q=\sigma, 0$ and $\sigma=\uparrow,\downarrow.$

The NDO constraint is an identity in terms of the Hubbard operators
\begin{equation}
X_i^{\uparrow\uparrow}+X_i^{\downarrow\downarrow}+X_i^{00}=1.
\label{1.3}\end{equation}
Those operators act in the $3d$ on-site Hilbert space
spanned by the spin up and spin down states $|\sigma\rangle$ and the vacuum state, $|0\rangle.$
The doubly occupied states are excluded by construction.
On the other hand, the identification $\tilde c^{\dagger}_{i\sigma}=X^{\sigma 0}_i$ implies that
$$\sum_{\sigma}\tilde c^{\dagger}_{i\sigma}\tilde c_{i\sigma}+ \tilde c_{i\uparrow}\tilde c^{\dagger}_{i\uparrow}=1-n_{i\uparrow}n_{i\downarrow}.$$
Therefore the operators (\ref{1.2}) establish a faithful representations of the Hubbard algebra, provided
$n_{i\uparrow}n_{i\downarrow}=0.$

There are many other possible Hubbard algebra representations, e.g., the so-called slave particle ones. Within a standard slave-boson representation, for example,
$ X_i^{\sigma 0}$ is equated to $f^{\dagger}_{i\sigma}b_i,$ where $f_{i\sigma}$ stands for a fermion
spinful operator and $b_i$ denotes a boson operator that keeps track of the charge degrees of freedom.
However, relation (\ref{1.3}) is no longer an identity, and it instead enforces the local constraint
$\sum_{\sigma}f^{\dagger}_{i\sigma}f_{\sigma}+b^{\dagger}_ib_i=1$. In contrast to the representation
(\ref{1.2}) this constraint cannot be lifted to the action through a global large coupling as it was be done
with the $U$-term in (\ref{1.0}).

Among several possible Hubbard algebra representations there is one that seems to be most appropriate
to address the physics close to half-filling.  In that regime, there are localized lattice electrons that carry spin degrees of freedom as well as a small number of conduction fermions -- dopons -- which are essentially vacancies hopping in the lattice.
Exactly the same degrees of freedom are relevant for a lattice Kondo model, which suggests that there may be some close relationship  between the two models. A starting point is therefore to establish a representation of the Hubbard
operators in terms of these localized and itinerant degrees of freedom. Such a representation was proposed in \cite{wen},
\begin{eqnarray}
\tilde c_{i}^{\dagger}
=\frac{1}{\sqrt{2}}(\frac{1}{2}-2\vec S_i\vec\tau)\tilde d_i,
\label{1.4}\end{eqnarray}
where $\vec\tau$ is the set of the Pauli matrices normalized by the condition $\vec\tau^2=3/4$.
In this framework, the localized electron is represented by the lattice spin $\vec S\in su(2)$
whereas  the mobile doped hole (dopon) is described by the projected hole operator,
$\tilde{d}_{i\sigma}=d_{i\sigma}(1-n^d_{i-\sigma})$.
Here $\tilde c^{\dagger}=(\tilde c^{\dagger}_{\uparrow},\tilde c^{\dagger}_{\downarrow})^{t}$ and
$\tilde d=(\tilde d_{\uparrow}, \tilde d_{\downarrow})^{t}$. The local dopon number operator is denoted
as $n^d_{i\sigma}.$
In principle, the "tilde" sign over the dopon
operators can be dropped, as that would have caused an error of order $O(n_d^2)$ which is not important
in the underdoped regime.

To accommodate these new operators one obviously needs to enlarge
the original on-site Hilbert space of quantum states. This enlarged
space is characterized by the state vectors $|\sigma a\rangle$ with
$\sigma=\Uparrow,\Downarrow$ labeling the spin projection of the
lattice spins and $a=0,\uparrow,\downarrow$ labeling the dopon
states (dopon double occupancy is not allowed). In this way the enlarged
Hilbert space becomes
\begin{equation}
{\cal H}^{enl}=\{|\Uparrow 0\rangle_i,|\Downarrow
0\rangle_i,|\Uparrow \downarrow\rangle_i, |\Downarrow
\uparrow\rangle_i,|\Uparrow \uparrow\rangle_i,|\Downarrow
\downarrow\rangle_i\},
\end{equation}
while in the original Hilbert space we can either have one electron
with spin $\sigma$ or a vacancy:
\begin{equation} {\cal H} =\{|\uparrow \rangle_i,|\downarrow
\rangle_i,|0\rangle_i\}\label{3v},
\end{equation}
The following mapping between the two spaces is then defined:
\begin{equation}
|\uparrow \rangle_i \leftrightarrow |\Uparrow 0\rangle_i, \quad
|\downarrow \rangle_i \leftrightarrow |\Downarrow 0\rangle_i,
\label{vacancy}\end{equation}
\begin{equation} |0 \rangle_i \leftrightarrow \frac{|\Uparrow \downarrow\rangle_i
- |\Downarrow \uparrow\rangle_i}{\sqrt{2}}\label{vacancy1}.
\end{equation}
The remaining triplet states in the enlarged Hilbert space, $\left(|\Uparrow
\downarrow\rangle_i + |\Downarrow
\uparrow\rangle_i\right)/\sqrt{2}$, $|\Uparrow \uparrow\rangle_i$,
$|\Downarrow \downarrow\rangle_i,$ are unphysical and should
therefore be removed in actual calculations. In this mapping, a
vacancy in the electronic system corresponds to a Kondo singlet formed by a
lattice spin and a dopon whereas the presence of an electron is
related to the absence of a dopon.

Although the operators (\ref{1.4}) fulfill the commutation relations of the Hubbard algebra they are not identical to the standard Hubbard operators. To become so they must span the lowest $3d$ fundamental representation
of the algebra fixed by Eq. (\ref{1.3}).
In terms of the $\vec S_i$ and $d_i$ operators, Eq.(\ref{1.3}) reads \cite{pfk}
\begin{eqnarray}
\vec{S_i} \cdot
\vec{s_i}+\frac{3}{4}n^d_i=0.
\label{1.5}
\end{eqnarray}
Here $\vec s_i=\sum_{\sigma,\sigma'}{d}_{i\sigma}^{\dagger}\vec\tau_{\sigma\sigma'}{d}_{i\sigma'}$
stands for a spin dopon operator that interacts with the local spin operator through the Kondo-type coupling. It is important that the operator on the left hand side of Eq. (\ref{1.5}) possesses only non-negative eigenvalues, which implies that the constraint (\ref{1.5}) can be imposed
by adding to the Hamiltonian the term $\lambda\sum_i(\vec{S_i}\cdot
\vec{s_i}+\frac{3}{4}n^d_i),\,\, \lambda\to +\infty.$
Physically the NDO constraint in the spin-dopon representation acquires  quite a nontrivial content:
the physical degrees of freedom are represented by the localized lattice
spin-up and spin -down states and a spin-dopon singlet state. Unphysical states are given by spin-dopon triplet states.

To establish an explicit relationship between the Kondo and Hubbard models,
consider a Kondo lattice  model at strong coupling $\lambda$,
\begin{eqnarray}
H_{\lambda}\,=\, 2t\sum_{ij\sigma}
{d}_{i\sigma}^{\dagger} {d}_{j\sigma}
+ \lambda
\sum_i(\vec{S_i} \cdot
\vec{s_i}+\frac{3}{4}n^d_i),
\label{1.6}
\end{eqnarray}
where $\lambda/t \to\infty$ and the $n^d_i$ term ensures that the theory remains finite as $\lambda\to +\infty.$

On the other hand, let us consider a canonical Kondo model with coupling $K$,
\begin{equation}
H_{K}= -t\sum_{ij\sigma}
{d}_{i\sigma}^{\dagger} {d}_{j\sigma}
+ K
\sum_i\vec{S_i} \cdot
\vec{s_i}.
\label{1.7}\end{equation}
In the limit $K\gg t$, it takes the form \cite{hirsch}.
\begin{equation}
H_{K}= \frac{t}{2}\sum_{ij\sigma}
\tilde {c}_{i\sigma}^{\dagger} \tilde {c}_{j\sigma}
- \frac{3K}{4}\sum_i(1-\tilde n_i) +{\cal O}(t^2/K),
\label{1.8}\end{equation}
where $\tilde n_i=\sum_{\sigma}\tilde {c}_{i\sigma}^{\dagger}\tilde {c}_{i\sigma}=1-\tilde n^d_i.$
Comparing this with Eq.(\ref{1.6}) gives
\begin{equation}
H_{\lambda= \infty}=
-t\sum_{ij,\sigma}\tilde c^{\dagger}_{i\sigma}\tilde c_{j\sigma}=H_{U=\infty},
\label{1.9}\end{equation}
We thus see that both the standard  Hubbard model (\ref{1.0}) and the Kondo model (\ref{1.6}) possess the same strong - coupling limit given by Eq.(\ref{1.9}).

However, Eq.(\ref{1.9}) should be considered as just a conjecture.
To prove it one needs to derive both the leading as well as the next-to-leading terms in an effective Hamiltonian.
For the reader's convenience and as a check against some well known results,
we first perform a relevant expansion of the standard Hubbard model (\ref{1.0}) in inverse powers of $U$.
To this end, we rewrite it in the following way,
\begin{equation}
H_{U} = H_0 + V,
\label{3.0} \end{equation}
with
\begin{equation}
H_0= U\sum_i
n_{i\uparrow}n_{i\downarrow},
\label{3.1} \end{equation}
\begin{equation}
V=-t\sum_{ij\sigma}
c_{i\sigma}^{\dagger} c_{j\sigma}.
\label{3.2}\end{equation}
In the large-$U$ limit, we can treat $V$ as a small perturbation.
The ground state of $H_0$ contains no doubly occupied sites and it is spanned by
the physical states given by the set (\ref{3v}):
$$|O_g\rangle =\{|\uparrow \rangle_i,|\downarrow
\rangle_i,|0\rangle_i\}.$$

Let us now define the operator $P=\prod_i P_i$ that projects into the ground state of $H_0:$
$$P_i=|0_i\rangle\langle 0_i|+|\uparrow_i\rangle\langle \uparrow_i|+|\downarrow_i\rangle\langle \downarrow_i|.$$
Up to second order in $V$, we can define the effective Hamiltonian:\cite{messiah}
\begin{equation}
H_{eff} = PVP+\sum_{{\phi_n}\ne O_g}\frac{PV|\phi_n\rangle\langle \phi_n|VP}{\epsilon_0-\epsilon_n},
\label{3.3}\end{equation}
where $\epsilon_0=0$ is the ground state energy and $|\phi_n\rangle$ is an eigenstate of $H_0$ with eigenvalue
$\epsilon_n.$
Here we have $Pc^{\dagger}_{\sigma}P=|\sigma\rangle\langle 0|=X^{\sigma 0}=\tilde c^{\dagger}_{\sigma}$ and
$Pc_{\sigma}P=|0\rangle\langle \sigma|=X^{0\sigma}=\tilde c_{\sigma},$ so that
\begin{equation}
PVP= -t\sum_{ij\sigma}
\tilde c_{i\sigma}^{\dagger} \tilde c_{j\sigma}.
\label{3.4}\end{equation}

For simplicity from now on we will be considering solely nearest neighbour sites $i$ and $j$ ignoring the $3$-site interactions. In such a case we get the following eigenstates of $H_0$ with eigenvalue $U:$
$$|\phi_1\rangle = |0\rangle_i|\uparrow,\downarrow\rangle_j$$ and $$|\phi_2\rangle=|0\rangle_j|\uparrow,\downarrow\rangle_i.$$
We have
$$PV|\phi_1\rangle= |\uparrow\rangle_i|\downarrow\rangle_j-|\downarrow\rangle_i|\uparrow\rangle_j.$$ The state
$PV|\phi_2\rangle$ is given by the same expression with the change $i\leftrightarrow j.$
As a result, we get
$$\sum_{\phi_n\ne O_g}PV|\phi_n\rangle\langle \phi_n| VP=$$
$$ 2(X^{\uparrow\uparrow}_iX^{\downarrow\downarrow}_j-X^{\uparrow\downarrow}_iX^{\downarrow\uparrow}_j-
X^{\downarrow\uparrow}_iX^{\uparrow\downarrow}_j +X^{\downarrow\downarrow}_iX^{\uparrow\uparrow}_j).$$

Equation (\ref{3.3}) then yields the well-known representation for the so-called $t-J$ model Hamiltonian,
$H_{eff}=H_{t-J},$ where
\begin{equation}
H_{t-J}
=-t\sum_{ij\sigma} \tilde{c}_{i\sigma}^{\dagger}
\tilde{c}_{j\sigma}+ J\sum_{ij} (\vec Q_i \cdot \vec Q_j-\frac{\tilde n_i\tilde n_j}{4}),
\label {3.5}\end{equation}
with three-site interactions being ignored.
Here $\vec
Q_i=\sum_{\sigma,\sigma'}\tilde {c}_{i\sigma}^{\dagger}\vec\tau_{\sigma\sigma'}\tilde {c}_{i\sigma'}=
\sum_{\sigma,\sigma'}{c}_{i\sigma}^{\dagger}\vec\tau_{\sigma\sigma'}{c}_{i\sigma'}$
is the local electron spin operator.
The emergent spin exchange coupling $J=4t^2/U.$ At $U=\infty$ the double occupancy virtual process is totally prohibited so that $J=0$.

Let us now turn to the Kondo model (\ref{1.6}):
\begin{equation}
H_{\lambda} = H_0 + V,
\label{3.7} \end{equation}
with
\begin{equation}
H_0= \lambda
\sum_i(\vec{S_i} \cdot
\vec{s_i}+\frac{3}{4}n^d_i),
\label{3.8}\end{equation}
and
\begin{equation}
V= 2t\sum_{ij\sigma}
{d}_{i\sigma}^{\dagger} {d}_{j\sigma}
\label{3.9}\end{equation}
In the large-$\lambda$ limit, we treat $V$ as a perturbation to $H_0$.

The ground state is spanned by the states
$$|O_g\rangle =\{|\Uparrow 0 \rangle_i,|\Downarrow 0
\rangle_i,\frac{|\Uparrow \downarrow\rangle_i-|\Downarrow \uparrow\rangle_i}{\sqrt 2}\}.$$
The projector into the ground state of $H_0$ now reads
$$P_i=|vac\rangle_i\langle vac|_i +|\Uparrow 0\rangle_i\langle \Uparrow 0|_i+ |\Downarrow 0\rangle_i\langle \Downarrow 0|_i,$$ where the vacancy state is
$$| vac\rangle_i=\frac{|\Uparrow \downarrow\rangle_i
- |\Downarrow \uparrow\rangle_i}{\sqrt{2}}.$$
We further get
$$Pd_{\uparrow i}P=-\frac{|\Downarrow 0\rangle_i\langle vac|_i}{\sqrt 2},$$ which under the identification (\ref{vacancy},\ref{vacancy1}) becomes $-\tilde c^{\dagger}_{\downarrow i}/\sqrt 2.$ In the same way,
$$Pd_{\downarrow i}P=\frac{\tilde c^{\dagger}_{\uparrow i}}{\sqrt 2},\,
Pd^{\dagger}_{\uparrow i}P=-\frac{\tilde c_{\downarrow i}}{\sqrt 2},\,
Pd^{\dagger}_{\downarrow i}P=\frac{\tilde c_{\uparrow i}}{\sqrt 2}.$$
As a result,
$$PVP=-t\sum_{ij,\sigma}\tilde c^{\dagger}_{i\sigma}\tilde c_{j\sigma},$$
which proves (\ref{1.9}).

To evaluate a slight departure from the infinite $\lambda$  limit, one needs to take into account
$8$ eigenstates of $H_0$ with eigenvalue $\lambda$ that contribute to Eq.(\ref{3.3}).
These are
$$ |\phi_1\rangle = |\Downarrow 0\rangle_i \frac{|\Uparrow \downarrow\rangle_j+|\Downarrow \uparrow\rangle_j}
{\sqrt 2},$$
$$ |\phi_2\rangle = |\Uparrow 0\rangle_i \frac{|\Uparrow \downarrow\rangle_j+|\Downarrow \uparrow\rangle_j}
{\sqrt 2}$$
$$|\phi_3\rangle =|\Downarrow 0\rangle_i|\Uparrow\uparrow\rangle_j, \quad
 |\phi_4\rangle =|\Uparrow 0\rangle_i|\Downarrow\downarrow\rangle_j.$$
The other $4$ states can be obtained through the change $i\leftrightarrow j$.
Equation (\ref{3.3}) then reduces to $H_{eff}=H_{t-V},$ where
\begin{equation}
H_{t-V}=-t\sum_{ij,\sigma}\tilde c^{\dagger}_{i\sigma}\tilde c_{j\sigma} + V\sum_{ij} \tilde n_i\tilde n_j.
\label{3.10}\end{equation}
The inter-site coupling $V=3t^2/\lambda$ with $V/t\ll 1.$
Let us once again notice that in deriving (\ref{3.10}) three-site terms have been ignored.
The $t-V$ model  allows for
the following virtual process:
a dopon is transferred from a Kondo spin-dopon singlet at site $i$ to a spin state to form a virtual triplet state, and back to the Kondo singlet at a nearest-neighbour site $j$.
Since a spin-dopon singlet carries charge and no spin this process gives rise to the effective charge-charge exchange
interaction--the repulsion
between the nearest-neighbour states.
This is in a clear contrast to the $t-J$ model where the opposite spin singly-occupied configurations are connected by virtual excitations to and from a doubly occupied state  to produce the inter-site antiferromagnetic spin exchange coupling.
Within the $t-V$ model, the inter-site repulsion emerges even in the absence of a long-range Coulomb interaction.
It is just a manifestation of short-range strong electron correlations. We thus see that short-range strong electron correlations may give rise to both spin-spin as well as charge-charge inter-site couplings.

In conclusion,
although in the strong coupling limit, the Hubbard model (\ref{1.0}) and the Kondo-lattice model (\ref{1.6})
are identical to each other,
at large but finite couplings those models
exhibit different physics:
The spin-spin exchange $\sim J$ enhances the AF spin ordered phase, whereas
the inter-site repulsion $\sim V$ enhances the formation of a charge-ordered phase.
The fact that charge ordering is only manifest at a larger doping in the pseudogap phase is a strong indication 
that the Kondo physics plays an important role as we move away from the light doping regime.

\end{document}